\documentclass[10pt,a4paper]{article}

\usepackage{graphicx}
\usepackage[utf8]{inputenc} 
\usepackage[T1]{fontenc}    
\usepackage{hyperref}       
\usepackage{url}            
\usepackage{booktabs}       
\usepackage{amsfonts}       
\usepackage{nicefrac}       
\usepackage{microtype}      

\usepackage[margin=2.5cm]{geometry}

\title{Echo-Based Quantum Memory}

\author{G. T. Campbell, K. R. Ferguson, M. J. Sellars,\\
B. C. Buchler and P. K. Lam\\
Centre for Quantum Computation and Communication Technology\\
Research School of Physics and Engineering\\
The Australian National University\\
Canberra, ACT 2601, Australia}

\date{}
\begin{document}
\maketitle

\newcommand{\bra}[1]{\langle #1 \vert}
\newcommand{\ket}[1]{\vert #1 \rangle}
\newcommand{\inner}[2]{\langle #1 \vert #2 \rangle}
\newcommand{\VT}[1]{\ensuremath{{V_{T#1}}}}
\newbox\sectsavebox
\setbox\sectsavebox=\hbox{\boldmath\VT{xyz}}

\newcommand{\YSO}{Y$_2$SiO$_5$}

\section{Introduction}

Attenuation in optical fibers means that it is necessary to use quantum repeaters as an element in future communication networks. Without them, the rate at which photons can be successfully sent through a long fiber is too low to send complex states. To overcome the non-deterministic nature of successful transmission events, memories must be used to store photons that correspond to successful entanglement swapping operations \cite{bussieres2013prospective}. A number of approaches to achieving quantum memories for light have been developed \cite{lvovsky2009optical}. Here, we focus on memories that are based on photon echos.

In everyday situations, the absorption of a photon into a material is not a reversible process. Information that may be carried by the photon is dissipated into the environment through noisy loss channels, such as spontaneous emission or non-radiative decay. The goal of a photon echo memory is to engineer a system for which this is not the case. Instead, an absorbing material is controlled such that the absorption process for incoming photons can be reversed and that the emitted photons carry the same quantum state as those that were absorbed.

The key to reversing the absorption of light lies in the inhomogeneous broadening of an optical transition, where an ensemble of optical emitters do not share the same transition frequency, but instead occupy a spread of frequencies. When light is absorbed into the ensemble, the dipole moments of all of the emitters will initially be aligned in phase with optical electric field. However, the emitters will quickly evolve to be out of phase with each other due to their different frequencies. If the emitters can later be brought back into phase with each other, the individual dipole moments will again create an electric field that matches the originally absorbed light, and a photon echo will be emitted. Provided that the absorption process can be exactly reversed and the dissipative processes that act on the individual emitters are slow compared to the storage time, the echo can approach the amplitude of the input light and act as an efficient memory.

\section{Overview of Photon Echo Techniques}

The first photon echoes \cite{PhysRevLett.13.567,PhysRev.141.391}, observed in a ruby crystal, were analogous to spin-echoes in nuclear magnetic resonance experiments \cite{Hahn1950}. These echoes used a $\pi$-pulse to rotate the Bloch vectors of the emitters by 180$^\circ$ about the $x$-axis of the Bloch sphere at a time $\tau$ after an initial excitation by a $\pi/2$-pulse. The $\pi$-pulse reverses the effect of the inhomogeneous broadening such that the emitters rephase after an additional time $\tau$ and emit an echo. This echo technique, however, can't be used to create a noiseless and efficient memory for quantum states. If the initial excitation of the ensemble is weak, the $\pi$-pulse will create a population inversion which leads to amplified spontaneous emission, and therefore noise, when the echo is generated \cite{PhysRevA.79.053851}. Another challenge is that any ensemble that is optically thick enough to fully absorb incoming light will also re-absorb part of the echo.

To overcome the problems of population inversion and re-absorption, a number of photon echo schemes have been developed. Here we focus on three that have been successfully used to perform storage and on-demand recall of light beyond the classical limit. First, we consider gradient echo memory (GEM), which is one implementation of a set of protocols based on controlled reversible inhomogeneous broadening (CRIB). This technique relies on an applied inhomogeneous broadening that can be reversed to achieve an echo. Second, we consider atomic frequency comb (AFC) memories that use spectral engineering of an inhomogeneous linewidth to generate a passive echo. Finally, we consider rephased amplified spontaneous emission (RASE), which uses a population inversion to create a pair of photons, one of which is stored in the ensemble.

The initial proposal for a CRIB memory utilised the fact that the Doppler broadening in an atomic ensemble is opposite for counter-propagating optical pulses \cite{moiseev2001complete}. This technique requires temporally storing coherence on an allowed optical transition, placing severe time constraints on the control pulses. It was proposed by Moiseev et al. that CRIB could be introduced in a solid-state system by co-doping a crystal with two species of ions \cite{Moiseev2003}. The first ion is optically active and is where the light is stored. The inhomogeneous broadening of the first ion is reversed by changing the state of the second. This technique requires the state-dependent shift introduced by the second ion to dominate all other sources of inhomogeneous broadening. A suitable system where this is satisfied has yet to be identified. Finally, it was also proposed by Sangouard et al. that an external field gradient transverse to the propagation direction of the stored light could be utilised \cite{sangouard2007analysis}. It is only possible to achieve such a field gradient if the length of the interaction region is much shorter than its transverse dimensions. As a result, it is difficult to achieve the large optical depths required for efficient memory operation.

\subsection{Gradient Echo Memory}

\begin{figure}
   \centering
   \includegraphics[width=\columnwidth]{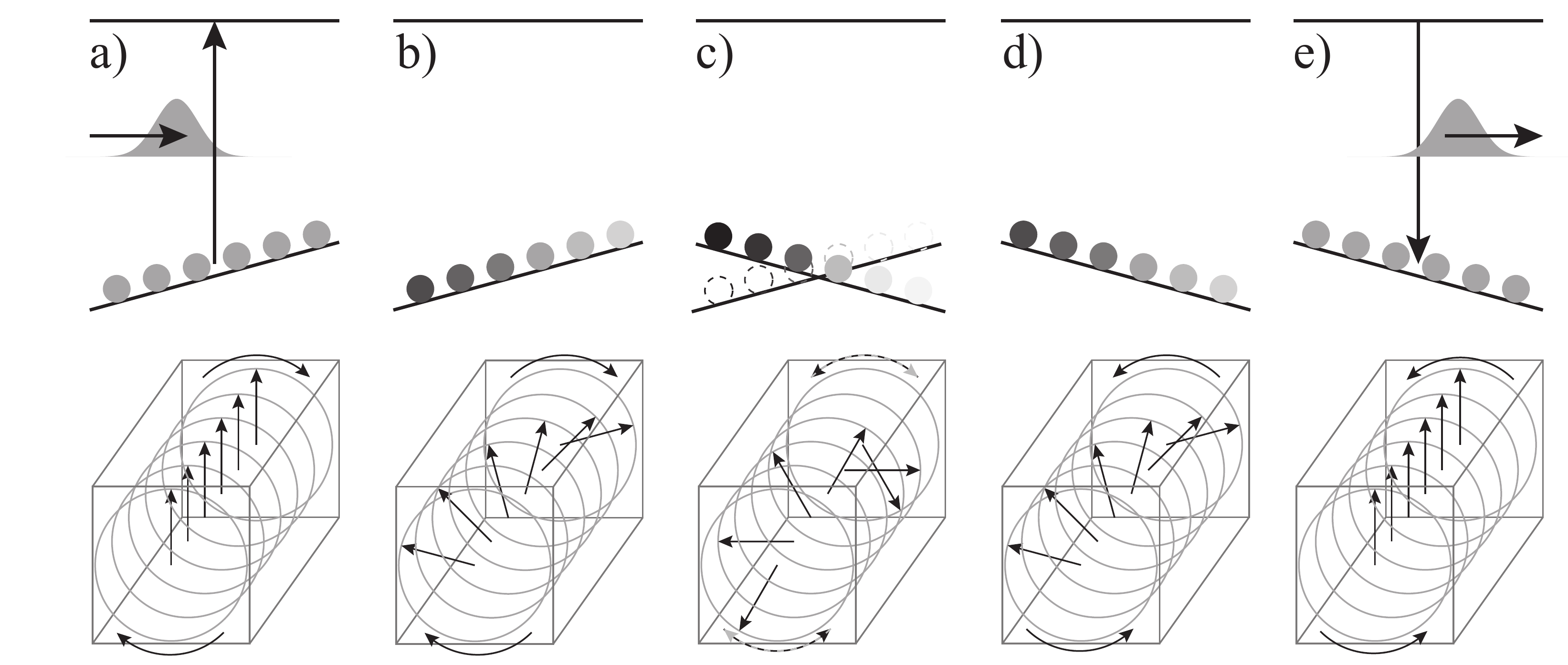}
   \caption{\label{fig:GEM}Atomic energy level structure (top) and projection of the Bloch vectors onto the $x-y$ plane (bottom) for GEM. A pulse is absorbed into a gradient-broadened ensemble (a). The atomic coherences then dephase (b) until the gradient is reversed (c). They then rephase (d) and emit an echo (e).}
 \end{figure}

A gradient echo memory uses the principle of CRIB, where the broadening is applied as a gradient along the propagation direction \cite{alexander2006photon,hetet2008electro}. The ensemble of emitters is taken to be a collection of two-level atoms that can be frequency-shifted by an external field. A spatial gradient can therefore be applied to the transition frequencies. An incoming optical pulse is fully absorbed if the ensemble has a large enough optical depth and if the bandwidth of the pulse is less than the broadening. The coherence created in the atoms by the absorption dephases as a result of broadening. An echo can then be obtained by reversing the sign of the gradient to rephase the coherence. The operation of the protocol is illustrated in fig.~\ref{fig:GEM}.

An important aspect of the protocol is that the gradient reversal avoids re-absorption of the echo. In other echo schemes, backward recall is required to achieve a time-reversal of absorption for efficient operation. For GEM, however, the broadening is applied spatially along the propagation direction and reversing it has the effect that the time reverse of absorption occurs in the forward direction. This results in a simple protocol because additional $\pi$-pulses are not required to reverse the propagation direction.

\subsection{Atomic Frequency Combs}

Another photon-echo technique relies on the engineering of a comb of narrow absorption lines within an inhomogenously broadened transition \cite{PhysRevA.79.052329}. Referred to as an atomic frequency comb (AFC) memory, the technique allows the ensemble to rephase on its own, without the need to apply refocusing pulses. The key is that the atoms are distributed into a set of absorption lines that are equally spaced in frequency. After an input optical pulse is absorbed into the comb of spectral features, each line of the comb will evolve in phase space. Because of the equal frequency spacing, all of the comb lines will return to the original phase at regular intervals.

\begin{figure}
  \centering
  \includegraphics[width=\columnwidth]{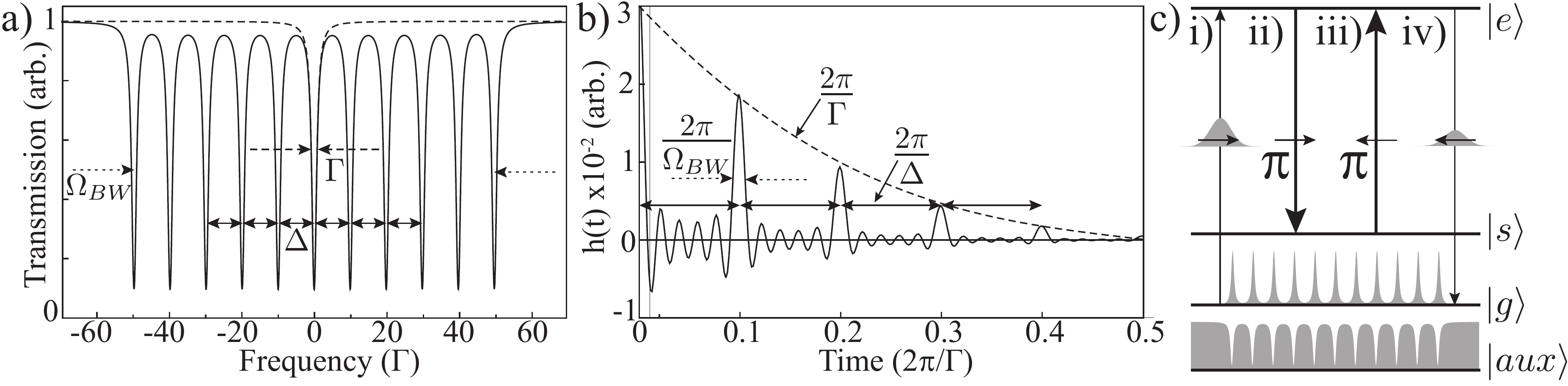}
  \caption{\label{fig:AFC}(a) The absorption profile of an atomic frequency comb. The periodic rephasing of the comb lines leads to a series of echoes in the corresponding impulse response (b). (c) An AFC is created by hole-burning selected atoms to an auxiliary state $\ket{aux}$. A pulse that is absorbed on the $\ket{g}\rightarrow\ket{e}$ transition (i) can be stored by transferring the coherence to a meta-stable state $\ket{s}$ using a $\pi$-pulse (ii). It can be recalled later using a second $\pi$-pulse in the backwards direction (iii) for an efficient echo (iv).}
 \end{figure}

Because the AFC echo arises from the spectral distribution of the ensemble, an example is shown in fig.~\ref{fig:AFC} (a), its properties can be understood by considering the impulse response associated with the absorption spectrum, illustrated in fig.~\ref{fig:AFC} (b). For a frequency spacing $\Delta$, echoes will occur at regular time intervals of $2\pi/\Delta$, within an envelope determined by the free-induction decay of a single comb tooth. For Lorentzian absorption lines of width $\Gamma$, this is an exponential decay with a time-constant $2\pi/\Gamma$. The shortest feature that can be efficiently stored is determined by the width of each echo in the response function $2\pi/\Omega_{BW}$, given by the overall width of the comb $\Omega_{BW}$. The delay-bandwidth product is then proportional to $\Omega_{BW}/\Delta \approx N$, where $N$ is the number of comb teeth. If a comb can be created with a large number of teeth, it can delay a large number of temporally separated pulses. 

The above description, however, is only half the picture. The absorption comb operates only as a delay-line and cannot recall a stored state on demand. Furthermore, a passive echo from an AFC, emitted in the forward direction, can be at most 54\% efficient due to re-absorption \cite{PhysRevA.79.052329}. To overcome this, the AFC protocol uses a $\pi$-pulse to store the excitation to a meta-stable state before the first echo is generated, illustrated in fig.~\ref{fig:AFC} (c). The excitation can then be transferred back to the excited state by a counter-propagating $\pi$-pulse, resulting in an echo in the backward direction. This exploits the time-reversal symmetry of the equations of motion \cite{moiseev2001complete} and can result in 100\% efficient, on-demand recall. 

The absorption comb structure of an AFC is produced by spectral hole-burning in an inhomogeneously broadened optical transition. This uses bright optical fields to selectively pump atoms that are resonant at certain frequencies to a dark state. An advantage of the AFC memory is that it rejects fewer of the ions from the inhomogeneous linewidth compared to creating a single narrow spectral feature. For delay-line implementations, this offers an extremely large bandwidth that is compatible with single photons from spontaneous parametric down-conversion or other high-bandwidth sources. However, implementing the complete on-demand memory protocol with a large bandwidth remains a challenge. The application of a $\pi$-pulse to store the excitation requires a meta-stable state that is separated in frequency from neighboring states by more than the bandwidth of the memory.

\subsection{Rephased Amplified Spontaneous Emission}

Whether the memory is to be used as a quantum repeater or for a source of on-demand single photons, the goal is to have a stored photon that is entangled with a traveling photon. The typical approach is to consider a source of photon pairs and store one of the produced photons in the memory. The technique of rephased amplified spontaneous emission (RASE) \cite{Ledingham2010} takes an alternate approach of generating a herald photon while storing an excitation that is entangled with it, effectively producing a pair of entangled photons with one already in memory similar to the DLCZ protocol \cite{Duan2001}.

\begin{figure}
  \centering
  \includegraphics[width=\columnwidth]{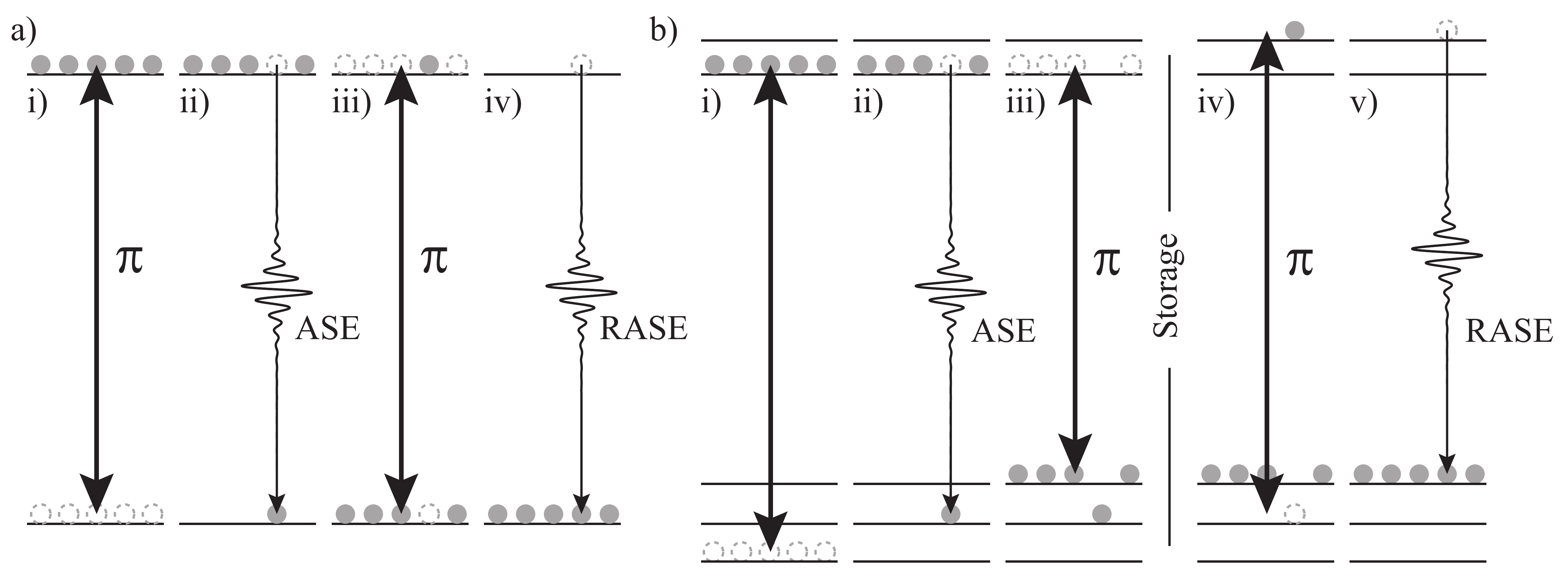}
  \caption{\label{fig:RASE}(a) Two-level RASE. A $\pi$-pulse inverts the population (i) leading to ASE (ii). A second $\pi$-pulse swaps the populations again (iii), resulting in RASE (iv). (b) The four-level RASE protocol adds a step to store the coherence on a meta-stable state using a $\pi$-pulse (iii). A rephasing $\pi$-pulse (iv) then creates a RASE echo (v). The ASE (ii) and RASE (v) fields are at different frequencies to all of the bright pulses.}
 \end{figure}

The RASE scheme, shown in fig.~\ref{fig:RASE} (a), works similarly to the two-level photon echo in reverse. A $\pi$-pulse is used to initially invert the population of an ensemble. Some of the atoms will then decay to the ground state via amplified spontaneous emission (ASE), leaving behind a coherence in the ensemble that is in phase with the emitted photon. Applying a $\pi$-pulse after an emission is detected returns most of the population to the ground state, but with an excitation that will rephase and create an echo of the emitted photon. The rephased amplified spontaneous emission (RASE) photon will be entangled with the initial ASE photon.

The approach can be extended by using a four-pulse echo sequence, shown in fig.~\ref{fig:RASE} (b), to shelve the excitation to a longer-lived spin state \cite{Beavan:11}. The four-pulse echo has the further advantage that each optical pulse has a unique frequency, simplifying the task of distinguishing the echo photon from free-induction decay of the $\pi$-pulses.

\section{Platforms for Echo-Based Quantum Memory}

The echo-based quantum memory protocols discussed above share a number of common requirements for any platform they are implemented in. First, long optical coherence times are required. This ensures that the ensemble excitation resulting from the absorbed input light remains in phase long enough for the read-out control fields to be applied. Second, while each of the echo techniques has a two-level variant, long-term storage can only be achieved if the coherence is transferred from the excited state to a state with a longer lifetime. The hyperfine splitting of the ground state is suitable, providing a manifold of states with long coherence times that can be used for three- and four-level protocols. However, addressing the hyperfine energy levels independently requires the hyperfine transitions to be resolved within the optical inhomogeneous line of the ensemble.

Once storage on the ground states has been achieved, dephasing processes must be minimized in order to maximize the attainable storage times. These dephasing processes arise from stochastic processes that apply to the entire ensemble and give rise to the homogeneous linewidth. In an atomic vapour dephasing mechanisms include atomic motion and collisions, while solid-state systems must contend with coupling to the crystal lattice and spin-flips in nearby atoms. 

A final requirement is that the optical depth must be sufficiently large to ensure a high probability of absorbing an input photon. In the GEM and AFC techniques the optical depth of interest is that of the broadened feature and the peaks of the teeth of the comb respectively. In the RASE scheme the situation is subtly different; high optical depth is required during the rephasing process to ensure high recall efficiency but low optical depth is required during the initial spontaneous emission event to ensure temporally well-resolved single photons \cite{Stevenson2014}. Switching the optical depth halfway through the sequence can be achieved in the four-level variant of the scheme by using transitions with different oscillator strengths for the initial and rephasing signals \cite{Beavan:11}.

No system that has currently been used to demonstrate an echo-based quantum memory has met each of these requirements intrinsically. However, there are techniques that can be applied to provide work-arounds in the most commonly employed platforms. In this section we detail the methods that must be employed to implement echo-based quantum memory techniques in two common platforms: rare-earth ion systems and vapors of alkali atoms. 

\subsection{Rare-earth Ion Systems}

A widely used platform for echo-based quantum memory demonstrations is rare-earth ions in either crystalline or amorphous materials. Commonly, the rare-earth ions are doped into a host material as impurities but stoichiometric crystals may offer compelling properties for quantum applications \cite{Ahlefeldt2016}. 

Rare-earth ion systems offer excellent properties for quantum memories because the valence electrons partially occupy the $4f$ orbital, which is shielded by the spatially larger $5s$ and $5p$ orbitals. This configuration renders rare-earth ions relatively insensitive to perturbation by the crystalline electric field. When cooled to cryogenic temperatures to reduce phonon broadening of the transition, extremely long optical coherence times are possible, as long as 4.38 ms \cite{Bottger2009}. In addition, ions exist with a wide range of wavelengths including some in the 1550 nm telecommunication band compatible with optical fibres \cite{Lauritzen2010}. Further advantages of rare-earth systems are an extremely large temporal \cite{PhysRevA.79.052329} and spatial \cite{heinze2013stopped} multimode capacity, and that the solid-state platform allows the creation of integrated waveguide architectures \cite{saglamyurek2011broadband,Marzban2015}.

While individual rare-earth ions can have extremely narrow optical transitions at low temperatures, the linewidth of an ensemble of ions is much larger, ranging from the order of one gigahertz to hundreds of gigahertz. The inhomogeneous broadening results from each ion experiencing a slightly different background electric and magnetic environment due to crystalline defects. The type and concentration of defects varies significantly for different materials.

To date, most quantum memory demonstrations in rare-earth ion doped crystals have used Pr$^{3+}$ or Eu$^{3+}$ doped in Y$_2$SiO$_5$ (YSO), because they are easy to work with. In these and similar crystals, the inhomogeneous broadening is much larger than the splittings of the hyperfine ground states, meaning that some method of resolving the hyperfine structure is needed for long-term storage of quantum states. Therefore, implementing on-demand memories with long storage times has relied on spectral hole-burning \cite{Pryde2000} to create a narrow absorption feature \cite{hedges2010efficient}, or a comb of narrow absorption features \cite{PhysRevLett.104.040503}, within a wide transparent trench that is created in the inhomogeneous linewidth.

To create the transparent trench, a laser is repeatedly swept over a large frequency range within the inhomogeneous linewidth. Ions are optically excited out of a level of interest and relax into a dark state where they can no longer interact with the laser. Most commonly the level of interest and the dark state are different hyperfine ground states. After many repetitions, all the ions are pumped into a dark state, creating a trench in the optical absorption spectrum. The maximum width of the transparent trench is limited by the separation between the level of interest and the dark storage level, i.e. the hyperfine splitting. The required absorption features can then be prepared inside the created trench by burning back a subset of the removed ions. Using this technique narrow features with controllable inhomogeneous broadening and optical depth can be engineered, allowing the feature properties to be tuned for the requirements of the specific memory protocol.

The disadvantage of the hole-burning technique is that it works by shelving the vast majority of the ions into non-interacting states, severely reducing the number of ions that can contribute to the effective optical depth. If instead a material can be engineered with an inhomogeneous broadening smaller than the hyperfine splitting, all the ions in the ensemble can be used in the operation of the memory, potentially offering much higher optical depths. 

Very narrow inhomogeneous linewidths have been observed in rare-earth ion doped crystals by using exceedingly low dopant concentrations and by isotopically purifying the host. Optical inhomogeneous lines as low as 16 MHz for Er$^{3+}$:Y$^7$LiF$_4$ \cite{Thiel2011} and 10 MHz for Nd$^{3+}$:Y$^7$LiF$_4$ \cite{Macfarlane1998} have been observed. However, while the hyperfine structure is resolved in these materials, the low dopant concentration results in a low optical depth. This means that little advantage is gained for quantum memory applications over engineering a subset of the ensemble from a crystal with a higher doping concentration. A promising option for achieving narrow optical inhomogeneous lines and high optical depths simultaneously is to use crystals that are stoichiometric in the rare-earth ion. Because of the concentration, optical depths in excess of 1000 cm$^{-1}$ are expected. Narrow linewidths have already been demonstrated in stoichiometric crystals, with a linewidth of 25 MHz obtained in EuCl$_3$.6H$_2$O by isotopically purifying the crystal in $^{35}$Cl \cite{Ahlefeldt2016}.  

An alternative is to use materials with larger hyperfine splittings, for example, in Er crystals. However, achieving long hyperfine coherence times in these materials has been more challenging. Recently, a hyperfine coherence time of 1.3 s has been demonstrated in $^{167}$Er$^{3+}$:YSO \cite{Rancic2016}. This material has ground state hyperfine splittings of the order of a GHz, larger than the optical inhomogeneous broadening. This material is an extremely promising candidate for demonstrating efficient quantum memories at telecommunication wavelengths. 

Once the hyperfine structure has been resolved, long-term storage of quantum states on the hyperfine ground states is possible. Several techniques can be employed to maximise the storage time. The major dephasing mechanism that limits the hyperfine coherence time at low temperatures is spin flips in the host crystal. These produce a fluctuating magnetic field that changes the transition energy. The zero first order Zeeman (ZEFOZ) technique uses a specific magnetic field to render the hyperfine transition insensitive to this perturbation. A second technique, dynamic decoherence control (DDC), uses a sequence of refocusing pulses to repeatedly flip the spin of the active rare-earth ion faster than the spin flips occur in the host crystal, averaging out the effect of the perturbation. DDC is only effective when inverting the spin of the active ion has negligible back-action on the spin-bath of the host crystal. This usually relies on the prior application of ZEFOZ to decouple the interaction between the spin of the active ion and the spin-bath. With these techniques, coherence times above one minute and above six hours have been achieved for Pr$^{3+}$:YSO \cite{heinze2013stopped} and Eu$^{3+}$:YSO \cite{zhong2015optically} respectively. These results open up the possibility of extremely long-lived quantum memories.

\subsection{Vapors of Alkali Atoms}

The other physical systems that have been used to demonstrate photon echo memories consist of vapors of alkali atoms.  The alkali atoms have large hyperfine splittings of the lowest energy level, in the GHz range, which is less than the inhomogenous linewidth. The resolvable hyperfine levels, with coherence times that can be in excess of a minute \cite{PhysRevLett.105.070801}, make an attractive candidate for storing quantum states. A further advantage of the alkali atoms is a relatively large absorption oscillator strength \cite{steck2001rubidium}. The hyperfine levels can also be split by the application of a magnetic field. The manifold of resolved magnetic sub-levels then provides a rich set of energy levels that can be used in memory protocols. While cesium has been used in slow light \cite{hsiao2016eit} and Raman \cite{PhysRevLett.116.090501} memory experiments, the only alkali atom that has been used for photon echo experiments is rubidium.

A challenge for using alkali atoms for photon echo memory, however, is the short lifetime of the optical transition, 28 ns for the $^{87}$Rb D$_1$ line for example \cite{steck2001rubidium}. To work around this, memories in alkali atoms use two-photon Raman transitions to drive transitions between hyperfine or magnetic sub-levels. For photon echo memories, these take the form of an off-resonant Raman transition formed by a bright optical coupling and the weaker signal field that is to be stored \cite{hosseini2012storage}. The off-resonant transition creates an effective two-level atom for the signal field. This effective transition has optical properties that can be controlled by the intensity and frequency of the coupling field as well as by applied magnetic fields.

A further challenge for alkali atom memories is the motion of the atoms through space. Warm vapor cells, operating just above room temperature, provide a convenient atomic ensemble but present the problem that the atoms are moving at thermal velocities. This leads to Doppler broadening of the excited state transition and transit broadening of the hyperfine transition. The transit broadening, resulting from atoms leaving the interaction region, can be reduced by the introduction of a buffer gas to slow the diffusion time. Alternatively, the alkali atoms can be trapped and cooled in a magneto-optical trap to sub-millikelvin temperatures.

Broadening mechanisms can be nearly eliminated in cold atomic ensembles. Coherence times, however, are limited to $\simeq$1 ms because the trapping fields must be switched off prior to operating the memory, resulting in the atomic ensemble falling out of the interaction region due to gravity. It is possible to overcome this limitation through the use of an optical lattice, which has been demonstrated to have a coherence time of 240 ms \cite{PhysRevLett.103.033003}.

\section{Characterization}

In order to integrate memories into quantum communication networks or optical computations, the performance of the memories must be well understood and meet the relevant requirements for a given application. While the requirements vary considerably depending on the proposed use, there are a number of classical and quantum criteria that can be used to quantify memory performance.

\subsection{Classical criteria}
\subsubsection{Efficiency}
 High efficiency makes the scaling of quantum networks much easier as it increases the success rate of entanglement distribution. In most memory schemes, photons are injected into the memory, stored and recalled.  If the memory is time symmetric, the efficiency of the storage process is equal to the efficiency of the recall process.  Finite optical depth, therefore, limits both the efficiency of trapping and recalling light \cite{PhysRevA.78.032337}.

 The RASE scheme differs from the others in that the stored photon is generated inside the memory. Whereas for GEM and AFC the total efficiency is defined as the ratio of the input and output photon numbers, in RASE, the efficiency is just the efficiency of the rephasing process that converts the atomic coherence into light \cite{Beavan2012,Ferguson2016}.  

\subsubsection{Bandwidth}
The bandwidth of a memory will determine the kind of photons that can be stored in the atoms.  One of the most successful sources of single and entangled photons, is spontaneous parametric downconversion (SPDC) \cite{Kwiat:1995ck}.  Unfortunately it is also a source with bandwidth of many THz unless it is somehow filtered or built within a cavity, which generally leads to a loss in brightness \cite{Bao:2008hh,Scholz:2009bd,Haase:2009ez,Rambach:2016iw,Rielander2016}.  A challenge for the practical application of quantum memory is therefore finding a route to compatible photon and memory bandwidth. One approach is protocols such as RASE, where the entangled photons are both generated and stored in a single system, ensuring perfect compatibility of the source and memory. 

\subsubsection{Storage time}  
In a fibre network signals travel 200~m$\mu s^{-1}$.  If the travel time between quantum repeater nodes is longer than the storage time, the quantum repeater will fail. A $1/e$ storage time greater than a millisecond is generally considered desirable for practical repeater applications, although some protocols have been developed that may relax this requirement \cite{Simon:2007dc,Matt45}. 

The product of the storage time and bandwidth, sometimes referred to as the delay-bandwidth product, is also a useful figure of merit because it reveals how many modes may be stored in the memory simultaneously \cite{lvovsky2009optical}. A large multimode capacity is required for some repeater protocols \cite{Simon:2007dc}.

\subsection{Quantum criteria}
The quantum state emerging from the memory will be preserved if no photons are lost and no noise is added. The efficiency deals with the loss of photons, but determining the added noise requires statistical analysis of the memory output.

\subsubsection{Fidelity}  
The fidelity is  the wavefunction overlap between the input and output states. Experiments are often characterized by comparing measured fidelity to the no-cloning limit. If the fidelity is below the no-cloning limit, an eavesdropper could, theoretically, collect enough information about the state to reconstruct a better version of the input than is available at the output of the memory.

Comparison of fidelity numbers between experiments is, however, fraught with difficulty for three main reasons. The first issue is that the no-cloning limit depends on the input state. For example, the coherent state limit is $2/3$ \cite{Cerf:2000fg}, for a single photon state it is $5/6$ \cite{Bruss:1998dw} and for thermal and squeezed states it will depend on the amount of noise and squeezing \cite{Olivares:2006id}. Comparing fidelities directly is therefore only possible if experiments use identical input states.

The second problem is that it makes some implicit assumptions about the efficiency of the system.  Consider the example of storing a very small coherent state ($\alpha\ll 1$). Provided the efficiency of the memory is perfect, the $2/3$ fidelity limit is a reasonable benchmark, but it is possible to cheat by having a memory that is actually a perfect absorber.  In this case the output of the memory will be a vacuum state, which can have overlap with the input state much greater than the cloning limit. A beam dump is clearly a poor quantum memory, but this is not captured by the fidelity.

The third pitfall is that the use of fidelity is different depending on whether the experiment is conditional or unconditional. Working with single photons provides the ability to measure conditionally. This means that only the times when a photon is output from the memory are considered. Thus, even when the memory efficiency is small, it is possible to measure very good quantum statistics by ignoring all the times the photon was lost by the memory. Conditioning can be achieved through measurements of single photons in a basis that is embedded in the photon, such as polarization \cite{Zhou:2012hn,Gundogan:2012iz,Clausen:2012gw}, timing \cite{sinclair2014spectral} or orbital angular momentum \cite{Nicolas:2014fna}. For this reason one must be careful when comparing fidelities of quantum memory experiments since some fidelities may be conditional, and others may not.  

\subsubsection{State independent metrics}
It is possible to come up with metrics that do not depend on the input state. One way is to simply plot the noise and compare this to the noise that would be introduced by an ideal quantum cloning device \cite{hedges2010efficient}. Another is to combine efficiency and noise data and plot results in terms of the signal-transfer (T) and conditional variance (V) to make a T-V diagram \cite{Hetet:2008dm,Hosseini:Nphys:2011}. In either case, the metrics can be used to verify how a memory is performing relative to the cloning limit.

Another possibility is to infer the behavior of the memory over all possible input states. One approach is system tomography \cite{Lobino:2009is}.  Another is to synthesize a virtual entangled state from an ensemble of mixed-state data and show preservation of entanglement \cite{Killoran:2012vra}. These approaches, while complete, have not been widely implemented.

\subsubsection{Entanglement preservation}
A final method of confirming quantum behavior in a memory is to store part of an entangled state and show that it remains entangled after storage.  One can then apply normal entanglement criteria such as the  inseparability criterion \cite{Duan2000}, the EPR criterion \cite{Einstein1935}, Clauser-Horne-Shimony-Holt (CHSH) Bell inequality \cite{Clauser1969} or a Cauchy-Schwarz type inequality \cite{Mandel1995} to show that entanglement is preserved. This brute-force method requires a suitable quantum source of light, which, as discussed in the bandwidth section above, is not always available.

\section{Demonstrations}

\subsection{Gradient Echo Memory}

The first demonstrations of a gradient inversion used to store and recall light were done using cryogenically-cooled rare-earth ion-doped crystals.  Electrodes were used to provide an electric field gradient across the crystal and thus a reversible Stark shift.  The initial demonstration with Eu$^{3+}$:YSO had an efficiency of about one part per million \cite{alexander2006photon}, but it was quickly increased to 15\% in Pr$^{3+}$:YSO \cite{hetet2008electro}.  These experiments were both done using two-level GEM with the long optical coherence times available in rare-earth systems.

The technique was subsequently transferred to a three-level $\Lambda$-scheme.  By using a strong off-resonant coupling beam to couple two low-lying spin states, light was stored and recalled from a spin coherence in $^{87}$Rb \cite{hetet2008photon}.  The gradient in this case was provided by magnetic field coils, providing a reversible Zeeman gradient along the atomic ensemble. The storage efficiency was 1.5\%.

Since these initial experiments, GEM has set a series of performance records for unconditional quantum memory. With the rare-earth Pr$^{3+}$:YSO platform an efficiency of 69\% has been demonstrated along with noise measurements that show surpassing of the no-cloning limit \cite{hedges2010efficient}. The three-level GEM scheme has recorded efficiencies as high as 87\% both in warm \cite{Hosseini:2011ks} and laser-cooled \cite{Cho:2016ksa} platforms. Both these schemes have also shown performance beyond the no-cloning limit characterized using fidelity and T-V criteria \cite{Hosseini:Nphys:2011,Cho:2016ksa}.  In terms of storage time the best achieved so far is 1~ms using laser-cooled atoms \cite{Cho:2016ksa}, although spin coherences in rare-earth systems have now been measured up to times of six hours \cite{Zhong:2015bw}.

GEM has also been applied to show various coherent manipulations. Since the frequency content of the light is stored spatially along the length of the ensemble, it lends itself to spectral manipulation of the stored light \cite{Sparkes:2012hea}. These experiments showed that it is possible to frequency-shift the recalled light and multiplex different frequencies of light at both the storage and recall stages of the experiment.  Using the extra flexibility afforded by the control beam in the three-level GEM scheme, experiments have also shown time-multiplexing of pulses, thus allowing reordering of pulses stored at separate times \cite{Hosseini:2009p8466}.  Interference between pulses has also been shown in the time and frequency domain \cite{Campbell:2012ew}.  In the time domain this was extended to build the analogue of a Fabry-Perot cavity. In this system the round-trip time of the cavity was determined by the 12~$\mu s$ memory time, thus giving an effective cavity round-trip path length of 3.6km \cite{Pinel:2015fq}.

The spatial multimode nature of the memory has been demonstrated via image storage in warm atomic vapor \cite{Higginbottom:2012gl}.  The time-multiplexing capabilities were also used to store multiple images in the memory with separate recall times \cite{Glorieux:2012ka}. GEM has also been used to generate stationary light in laser-cooled atoms \cite{Everett:2016eb}. In this experiment the frequency-binning nature of the memory was used to create spinwaves capable of sustaining stationary light.

\subsection{AFC demonstrations}

The first demonstration of an atomic frequency comb echo was conducted in Nd$^{3+}$:YVO$_4$ \cite{de2008solid}. While the efficiency in this initial work was limited to 0.5\% by the method of comb preparation, a delay-bandwidth product of 25 was shown by delaying 20 ns pulses for 500 ns. Additionally, the memory was shown to be capable of storing multiple time-separated pulses. Further work, conducted using Pr$^{3+}$:YSO, improved the efficiency of a forward-recall AFC delay to 35\% \cite{amari2010towards}. The large multimode capacity of AFCs has also demonstrated by delaying 1060 temporal modes in Tm$^{3+}$:YAG \cite{Bonarota2011} and 26 spectral modes in Tm$^{3+}$:LiNbO$_3$ \cite{sinclair2014spectral}.

AFC delay-lines were shown to operate in the quantum regime with two experiments \cite{clausen2011quantum,saglamyurek2011broadband} that stored one of a pair of broadband entangled photons produced by SPDC. Both experiments demonstrated a violation of the CHSH Bell inequality between the stored photon and a herald photon at a wavelength suitable for fiber telecommunications. One experiment \cite{clausen2011quantum} was conducted in Nd$^{3+}$:YSO with an efficiency of 21\% and storage time of 25 ns. The other experiment \cite{saglamyurek2011broadband} was done using Tm$^{3+}$:LiNbO$_3$ with an efficiency of 2\% and storage time of 7 ns. Notably, the demonstration in Tm$^{3+}$:LiNbO$_3$ was in a waveguide, showing promise for integrating rare-earth memories into photonic circuits.

The first demonstration of an on-demand AFC memory was conducted in Pr$^{3+}$:YSO \cite{PhysRevLett.104.040503} with an efficiency of $\simeq$1\% and a storage time of $\simeq$10 $\mu$s. A technical challenge of the full AFC scheme is to filter the bright $\pi$-pulses sufficiently that noise, either due to scattering or fluorescence, is not added to the output state. It was later shown that it is possible to achieve such filtering in an on-demand AFC memory at the single photon level. This was first demonstrated, again in Pr$^{3+}$:YSO, \cite{PhysRevLett.114.230501} by storing two weak coherent states that were separated in time and using two successive write pulses as a beamsplitter \cite{PhysRevA.88.022324} to interfere them. By varying the relative phase of the pulses and measuring the output with a single photon counter, the fidelity of the storage was shown to surpass the classical limit. The efficiency in the experiment was 5.6\% with a storage time of 20 $\mu$s.

While the efficiencies of initial AFC demonstrations have been relatively low, techniques are being developed to improve them. A significant limiting factor is the optical depth of the ensembles. The use of optical cavities around the memory crystal \cite{sabooni2013cavity} has resulted in an AFC delay-line efficiency of 53\% and an efficiency with spin-wave storage for 10 $\mu$s of 28\% \cite{jobez2014cavity}. 

Work is also being done to implement AFC memories in platforms that can be integrated into optical networks. AFC delay-lines have been demonstrated in both erbium-doped fibers \cite{saglamyurek2015quantum} and Tm$^{3+}$:LiNbO$_3$ waveguides \cite{saglamyurek2011broadband}. A waveguide AFC memory with spin-wave storage has also been shown in Pr$^{3+}$:YSO by direct laser writing \cite{PhysRevApplied.5.054013}. In addition to paths to integrated photonic components, the waveguide demonstrations hold promise for non-linear optical gates due to the small mode volume of the guided light \cite{sinclair2016proposal}. 

\subsection{RASE demonstrations}

The major experimental challenge that must be overcome to demonstrate a quantum memory using the RASE scheme is to isolate the single photon signals from the free-induction decay of the $\pi$-pulses. Two initial experimental demonstrations, undertaken simultaneously, took different approaches to eliminate this added noise. The first implemented the basic two-level RASE scheme in Tm$^{3+}$:YAG and minimised the noise by using highly uniform driving pulses \cite{Ledingham2012}. They violated the inseparability criterion with 95\% confidence, showing continuous variable entanglement between the initial ASE and its echo. The second demonstration, conducted in Pr$^{3+}$:YSO, used the four-level rephasing sequence, which allowed the ASE and RASE photons to be spectrally and spatially distinguished from the noise \cite{Beavan2012}. Single photon detection was used and, while a clear correlation was shown between the ASE and RASE intensities, there was an insufficient signal-to-noise ratio to show that the correlation was non-classical. 

Both of these initial demonstrations were limited by added noise. The two-level scheme is intrinsically noisy as the free-induction decay of the rephasing $\pi$-pulse is indistinguishable from the output of the memory. However, the four-level scheme has the potential for low noise operation with improved spectral filtering. A subsequent experiment conducted in Pr$^{3+}$:YSO used heterodyne detection to remove the need for complex frequency filtering and demonstrated that low noise operation was achievable in the four-level RASE scheme. They showed continuous variable entanglement between the ASE and the rephased field by violating the inseparability criterion with 98.6\% confidence \cite{Ferguson2016}.  In addition, the entanglement was preserved after storage of the coherence on the hyperfine ground states for up to 5~$\mu$s and the multimode capability of the RASE scheme demonstrated with the storage of two temporal modes. The main factor limiting the performance of RASE as a quantum memory is the low recall efficiency, with an efficiency of 3\% achieved in the four-level scheme \cite{Ferguson2016}. However, the efficiency is theoretically predicted to approach 100\% by using an impedance-matched optical cavity \cite{Williamson2014}.

\section{Outlook}

Each of the echo-based memory protocols discussed above has excelled in demonstrating a different performance characteristic.
GEM has succeeded in demonstrating extremely high efficiencies due to the advantage of having no re-absorption for recall in the forward direction. This negates the need to apply counter-propagating $\pi$-pulses. The AFC delay-lines have made good use of materials with large inhomogeneous broadening to achieve high bandwidths and large multimode capability. Finally, RASE is currently the only protocol with on-demand retrieval to have successfully stored a quantum state, preserving the entanglement. This is because generating and storing the entanglement in a single protocol ensures that the memory and source are perfectly compatible, which is otherwise challenging to achieve. 

While currently no protocol has been able to demonstrate all of the desired characteristics, all of the protocols have the capacity for significant improvement. The current limitations are largely caused by the properties of the systems the memories are implemented in, rather than any intrinsic constraints imposed by the protocols themselves. 

To build better quantum memories, new materials must be developed. An ideal material would have larger hyperfine splittings to allow large bandwidths, long hyperfine coherence times to enable long-term storage of the quantum information, and low disorder to allow all the available emitters to be used for the memory operation and provide a large optical depths.

Some promising materials have recently been demonstrated. In particular, $^{167}$Er$^{3+}$ in YSO has already shown a hyperfine coherence time of 1.3 s without using ZEFOZ \cite{Rancic2016}. This material will allow potentially GHz-bandwidth spin-wave storage, which could drastically improve the performance of the echo-based memory protocols. In addition, quantum memories in $^{167}$Er$^{3+}$:YSO would operate in the 1550 nm telecommunication band, allowing them to be easily integrated into the existing fiber communication network. Other promising options could include stoichiometric rare-earth crystals \cite{Ahlefeldt2016}, or optical lattices of alkali vapors \cite{PhysRevLett.103.033003}.

\bibliography{EchoMemoryBib}
\bibliographystyle{ieeetr}

\end{document}